\documentclass[11pt]{article}
\usepackage[sort&compress,square,comma,numbers]{natbib}

\usepackage[labelfont=bf]{caption} 
\usepackage[font={small}]{caption} 

\usepackage{subcaption}
\usepackage{amsmath}
\usepackage{amssymb}
\usepackage{url}
\usepackage{placeins}
\usepackage{textcomp}
\usepackage{graphicx}
\usepackage{float} 
\usepackage{chngpage}
\usepackage{authblk}

\usepackage[usenames,dvipsnames]{color}

\usepackage{lineno}

\newcommand*\patchAmsMathEnvironmentForLineno[1]{%
  \expandafter\let\csname old#1\expandafter\endcsname\csname #1\endcsname
  \expandafter\let\csname oldend#1\expandafter\endcsname\csname end#1\endcsname
  \renewenvironment{#1}%
     {\linenomath\csname old#1\endcsname}%
     {\csname oldend#1\endcsname\endlinenomath}}%
\newcommand*\patchBothAmsMathEnvironmentsForLineno[1]{%
  \patchAmsMathEnvironmentForLineno{#1}%
  \patchAmsMathEnvironmentForLineno{#1*}}%
\AtBeginDocument{%
\patchBothAmsMathEnvironmentsForLineno{equation}%
\patchBothAmsMathEnvironmentsForLineno{align}%
\patchBothAmsMathEnvironmentsForLineno{flalign}%
\patchBothAmsMathEnvironmentsForLineno{alignat}%
\patchBothAmsMathEnvironmentsForLineno{gather}%
\patchBothAmsMathEnvironmentsForLineno{multline}%
}

\usepackage[usenames,dvipsnames]{color}    
\usepackage{soul}   
\setstcolor{red} 
\setul{}{1.5pt}

\usepackage{pifont}

\newcommand{\smallequal}{\scalebox{0.75}[1.0]{\( = \)}}
\newcommand{\smallminus}{\scalebox{0.75}[1.0]{\( - \)}}
\newcommand{\smallplus}{\scalebox{0.75}[0.75]{\( + \)}}
\newcommand{\smallcos}{\scalebox{0.9}[0.9]{\( \hspace{0.1cm}\cos \hspace{0.1cm}k_y\)}}

\begin{document}


\title{\textbf{Trapped state at a dislocation in a weak magnetomechanical topological
   insulator}}

\author{Inbar Hotzen Grinberg$^1$, Mao Lin$^2$,
	\mbox{Wladimir A. Benalcazar$^2$},
	\mbox{Taylor L. Hughes$^{2\ast}$}, and Gaurav Bahl$^{1\ast}$\\
	\footnotesize{$^1$ Department of Mechanical Science and Engineering, $^2$ Department of Physics,}\\
	\footnotesize{University of Illinois at Urbana-Champaign, Urbana, Illinois 61801, USA}\\
	\footnotesize{$^\ast$ To whom correspondence should be addressed; hughest@illinois.edu, bahl@illinois.edu} 
}

\date{}

	\vspace*{-2cm}
	{\let\newpage\relax\maketitle}
	
	\begin{abstract}

Topological insulators (TIs) are characterized by an insulating bulk and symmetry protected bound states on their boundaries. A ``strong'' topological insulator is characterized by robust conducting states on \emph{all} boundaries protected by certain internal symmetries. A ``weak'' topological insulator (WTI) however, requires lattice translation symmetry, making it more sensitive to disorder. However, this sensitivity gives rise to interesting characteristics such as anisotropic edge modes, quantized charge polarization, and bound states appearing at dislocation defects. 
Despite hosting interesting features, the sensitivity of WTIs to disorder poses an experimental confirmation challenge. Here we realize a 2D magneto-mechanical metamaterial and demonstrate experimentally the unique features of a WTI. Specifically, we show that the 2D WTI is anisotropic and hosts edge modes only on certain edges, as well as hosting a bound state at a dislocation defect. We construct the 2D WTI from stacked 1D Su-Schrieffer-Heeger (SSH) chains for which we show experimentally the different gapped phases of the 1D model.

\end{abstract}

\section{Introduction}

Topological insulators (TIs) are bulk-insulating materials with symmetry protected conducting states on their boundaries \cite{hasan2010colloquium,qi2011topological,moore2010birth,bernevig2013topological}. 
Since its discovery \cite{bernevig2006quantum,konig2007quantum}, the concept of TIs has expanded and have been demonstrated in different metamaterial systems, including photonic crystals \cite{raghu2008analogs,haldane2008possible,lu2014topological,ozawa2019topological}, 
mechanical systems {\cite{kane2014topological,huber2016topological,bertoldi2017flexible,nash2015topological,mitchell2018amorphous}} and acoustic systems \cite{yang2015topological,zhang2018topological}. 
A ``strong'' TI is characterized by a quantized topological invariant and hosts robust conducting states on all boundary terminations. Furthermore, the boundary states, and any observable properties dependent on the topological invariant \cite{xiao2010berry,qi2008}, are protected against disorder and defects by certain discrete, internal symmetries \cite{kitaev2009,ryufurusaki2008,chiu2016classification,ryu2010topological}. Soon after the initial predictions of time-reversal invariant topological insulators, the concept of ``weak" topological insulators (WTIs) was proposed \cite{fu2007topological,fu2007topological2,noguchi2019weak}. In this context the WTIs were protected only in the presence of time-reversal symmetry and lattice translation symmetry, which makes them, in principle, more sensitive to disorder, and hence weaker. The possible classes of WTIs were extended to allow for many different types of internal symmetries, but they are all linked by the requirement for some type of lattice translation symmetry \cite{kitaev2009}. 

Strong TIs are characterized by quantized electromagnetic response properties which are isotropic and manifest as robust boundary states {on} all boundaries \cite{qi2008}. In contrast, since the symmetry protection of WTI{s} is related to lattice translation symmetry, WTIs are typically anisotropic, and exhibit low-energy boundary modes only on certain boundary terminations and orientations \cite{fu2007topological,fu2007topological2,noguchi2019weak}.
WTIs may also exhibit interesting anisotropic electromagnetic properties, e.g. quantized charge polarization \cite{ramamurthy2015patterns}, but again, these are only quantized in the presence of lattice translation symmetry. In addition to these two types of properties, Ref. \cite{ran2009one} made the remarkable prediction that crystal dislocations, which are essentially symmetry fluxes for lattice translation symmetry \cite{teo2010topological,teo2017topological,slager2014interplay,jurivcic2012universal,slager2018translational}, can trap mid-gap topological bound states that can be observed spectroscopically. 

\begin{figure}[t!]
		\vspace{0.5cm} 
		\makebox[\textwidth][c]{\includegraphics[width=1\textwidth]{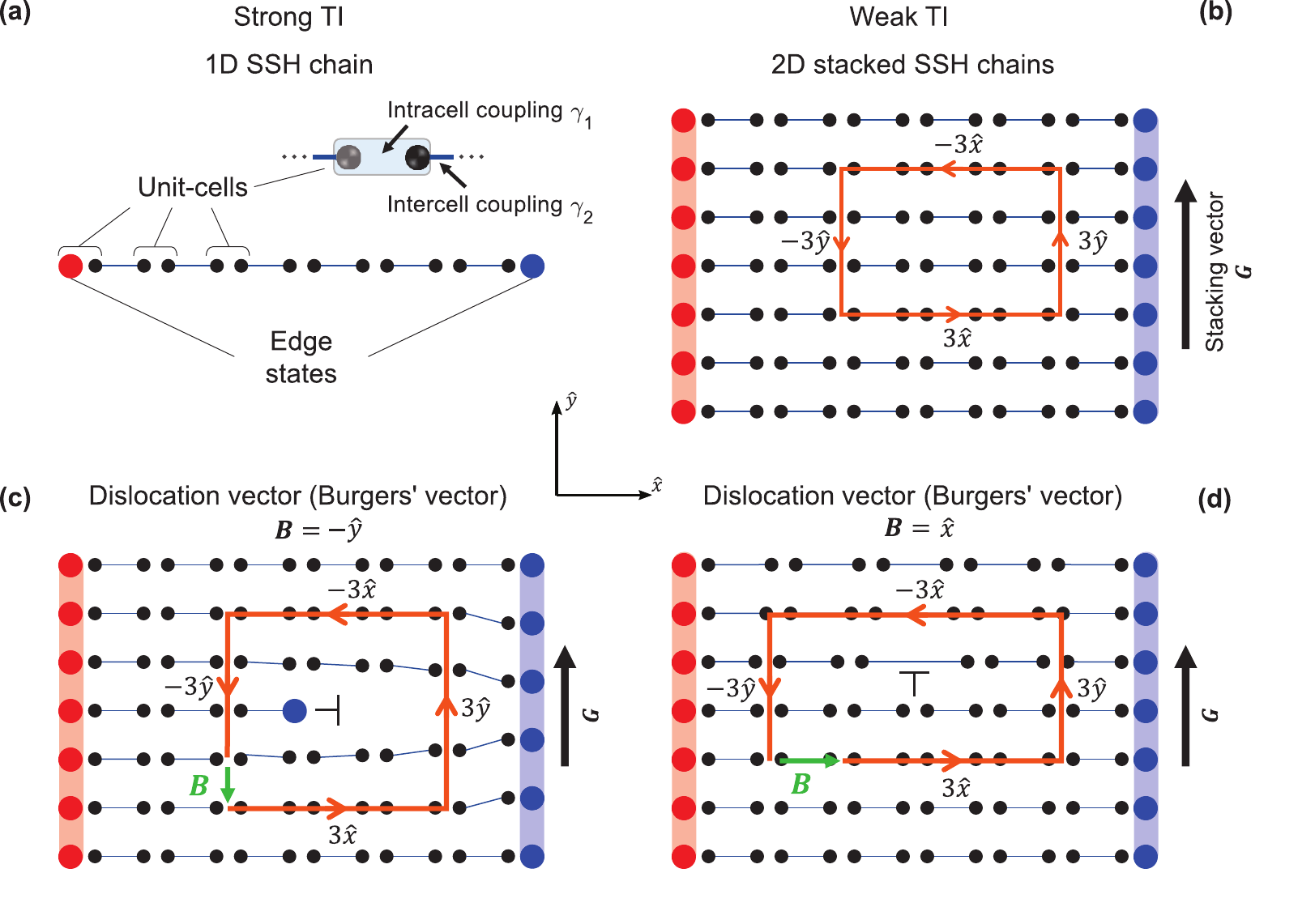}}
		\centering
		\caption{
			\textbf{(a)} Illustration of a dimerized  array forming a 1D topological insulator - the SSH chain. For visual simplicity we illustrate the limiting case where intra-cell coupling {$\gamma_1$ is zero}. This 1D array exhibits two mid-gap states localized on the two edges.
			\textbf{(b)} A 2D WTI is formed by stacking SSH 1D TIs while preserving translation symmetry. Localized mid-gap modes should appear  on edges that are parallel to the stacking vector ${\bf G}$. 
			In this array, starting from any point and taking $N_\text{step}$ ($N_\text{step}$ is an integer) steps in each direction (the step size is equal to the lattice constant for both directions) will always result in a closed path. Here we demonstrate the case with $N_\text{step}=3$, and the closed path is indicated by orange arrows. Taking the same closed path around a dislocation will result in a mismatch between the beginning and end points. The Burgers vector {\bf B} is the vector connecting these two points and characterizes the dislocation.
			\textbf{(c)} A 2D WTI with a dislocation defect having Burgers' vector ${\bf B} = -\hat{y}$ traps a localized mode at the dislocation core.
			\textbf{(d)} A 2D WTI with a dislocation defect having Burgers' vector ${\bf B} = \hat{x}$, does not trap a mode at the dislocation core due to orthogonality with the stacking vector ${\bf G}$.		
		}
	\label{fig:Concept}
\end{figure}

Let us illustrate these properties in a simple limit. All WTIs can be generated by stacking lower-dimensional strong TIs into a periodic array \cite{fu2007topological}. As an example, we take copies of a strong 1D TI aligned along the $x$-direction (Fig.~\ref{fig:Concept}a) and stack them in the $y$-direction with equal spacing (Fig.~\ref{fig:Concept}b).
This construction generates a set of lattice lines, the 1D TI chains, parameterized by a reciprocal vector ${\bf{G}}$
 that is orthogonal to the chains, and sets a stacking direction (${\bf G} = (2\pi/a) \, \hat{v}$ for stacking in the direction of unit vector $\hat{v}$ in real space, where $a$ is the lattice constant).
 The resulting WTI is protected by whichever internal symmetries required to protect the 1D strong TI  as well as by lattice translation symmetry in the stacking direction. Although the WTI phase remains stable when the 1D chains are coupled in the stacking direction, we can identify any interesting topological properties in the limit when the chains remain decoupled. We note that in our experiments the coupling is always nonzero, however, for the sake of clarity we first illustrate these phenomena in {this} decoupled limit.
The 1D TIs have end states, {as a result of which} the resulting 2D WTI will have edge states comprised of the stacked 1D end states on edges parallel to the $y$-axis. However, it will not have edge states on edges parallel to the $x$-axis, which illustrates the surface anisotropy \cite{fu2007topological}. If the 1D TI has a quantized electromagnetic property then the WTI will typically exhibit the same property, but with a coefficient that depends on the number of layers in the ${\bf{G}}$ direction \cite{franz2013topological}.

To illustrate the dislocation bound state in a WTI we can introduce an edge dislocation{, which is essentially}  an extra partial line of sites into the system. If the partial line is parallel to the 1D TI chains (Fig.~\ref{fig:Concept}c), then {the termination of this extra line, i.e. the dislocation core,} will harbor a localized end-state from that 1D TI. The Burgers vector ${\bf{B}}$ of such a dislocation\footnote{A dislocation defect in the lattice is characterized by a mismatch between the beginning and end points when tracing a closed-loop of lattice translations around the dislocation core \cite{kittel1976introduction}. The vector connecting the end and start point is the Burgers vector ${\bf B}$ indicating the amount of translation in units of lattice constant.} will be parallel to ${\bf{G}},$ and hence, if we apply the topological index theorem introduced in Ref. \cite{ran2009one} we can count the number of stable dislocation modes to be  $n={\bf G}\cdot{\bf B}/2\pi\, \textrm{{modulo}\ }  2$.
For the example shown in Fig.~\ref{fig:Concept}(c) we have ${\bf G}=2\pi\hat{y}$ (setting lattice constant $a=1$) and ${\bf B}=-\hat{y}$, whereas in Fig.~\ref{fig:Concept}(d) {\bf G} remains the same but ${\bf B}=\hat{x}$.
Thus, if we insert an odd (even) number of partial lattice lines there will be one (zero) protected mid-gap modes trapped at the dislocation core.
 While the picture described above provides a simple description of the interesting WTI phenomena in the decoupled limit, the remarkable thing is that it survives even when the chains are (strongly) coupled as long as the bulk gap does not close and the symmetries are preserved. 
Thus we would find that the edge states that exist on the edges parallel to the $y$-axis will couple to each other and disperse along the edge, and the dislocation defect traps a 0D mode that is exponentially localized on the dislocation core in a 2D system. If ${\bf B}$ is orthogonal to ${\bf G}$, however, no mode will be trapped at the dislocation defect (Fig.~\ref{fig:Concept}d).

Despite having a range of interesting features, WTIs pose a challenge to confirm experimentally due to their fragility to disorder \cite{pauly2015subnanometre,hamasaki2017dislocation}, and as a result, WTIs have only been recently verified experimentally in solid state systems \cite{noguchi2019weak} and photonic crystals \cite{yang2019realization,li2018topological}. 
Here we take another approach for the realization of WTIs using a 2D magneto-mechanical metamaterial\cite{grinberg2019robust}. 
We exhibit the important spectroscopic features mentioned above, i.e., the anisotropic edge states, and the dislocation bound state.  
As the building block of our 2D WTI we use a 1D Su-Schrieffer-Heeger (SSH) strong TI \cite{Su1979,heeger1988solitons,su1980soliton} composed of a dimerized chain of mechanical resonators.
As a proof of concept, we first demonstrate different gapped phases of the SSH chain as well as a localized mode trapped on the domain wall between a topologically trivial and nontrivial phases. 
We then generate a WTI by stacking the SSH chains in the transverse direction.
We demonstrate the WTI anisotropy by showing the existence of edge modes in only one direction (i.e. left and right but not on top and bottom). Upon introducing a dislocation with Burger's vector orthogonal to the stacking direction, we observe a bound state trapped at the core of the dislocation which signals the nontrivial topology of the deformed lattice.

\vspace{12pt}


\section{Experimental results}

\begin{figure}[hp]
		
		\makebox[\textwidth][c]{\includegraphics[width=1.4\textwidth]{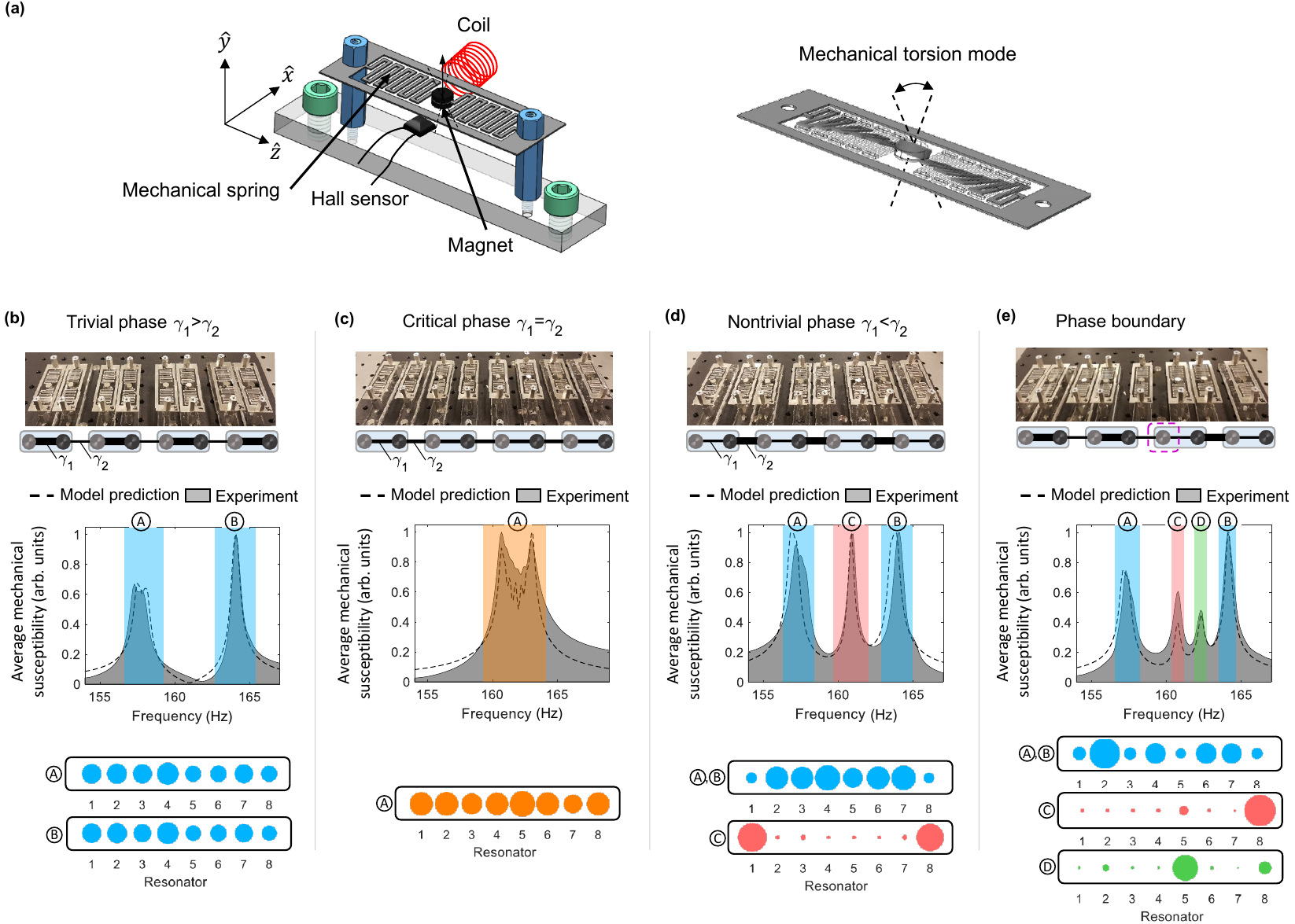}}
		\centering
		\caption{
			\textbf{(a)} Illustration of a magneto-mechanical resonator and its torsional resonance mode. The mechanical restoring torque is produced by an aluminum serpentine spring. A drive coil and Hall sensor are used to make local spectroscopic measurements at each resonator. We perform a series of experiments for 1D configurations of a dimerized resonator chain, corresponding to the  
			\textbf{(b)} trivial phase (insulator), 
			\textbf{(c)} critical point (metallic), 
			\textbf{(d)} and nontrivial phase (topological insulator). 
			\textbf{(e)} Demonstration of a domain wall between a trivial and a nontrivial phase.
			For each case in (b)-(e) we present a photograph of the experimental setup with a tight binding model illustration (thicker lines correspond to higher coupling rate). 
	We also present experimentally measured system-wide normalized mechanical susceptibility, which is the averaged system susceptibility at each frequency and corresponds to the mechanical density of states. The dashed lines are the theoretical model prediction. Figures also include the spatial distribution of states averaged over the highlighted bands (circle size corresponds to excitation amplitude). 
		}
		\label{fig:1dResults}
\end{figure}

We experimentally implement the 1D SSH chains using an array of identical mechanical resonators coupled through magnetic interaction. 
Each resonator has a single rotational degree of freedom $\theta$ around the $\hat{z}$ axis and is designed to operate in its torsional resonance mode (Fig. \ref{fig:1dResults}a). 
The resonance is facilitated by an aluminum serpentine spring that provides a mechanical restoring torque when $\theta \neq 0$. 
Each resonator has a neodymium magnet bonded to a central platform which serves as the resonating mass, while simultaneously producing a magnetic field through which adjacent resonators are coupled 
\cite{Supplement}.  
This magnetic field decays cubically with distance, allowing control of the coupling rates by manipulating the spacing between resonators. Each mechanical resonator is an analogous to an atom and we define a unit cell as two adjacent resonators with intra-cell coupling $\gamma_1$. We can then arrange these unit cells with periodicity along $\hat{x}$ {and} $\hat{y}$ to produce desired structures.

The magnetically induced torque between dipoles introduces an additional spring effect that can either soften or stiffen the torsional mechanical stiffness of each resonator \cite{grinberg2019magnetostatic} depending on its local magnetic environment. 
Since the resonators on the ends of these arrays have only a single neighbor, they experience a different magnetostatic spring effect compared against resonators within the bulk, and as a result their resonance frequency is detuned. We compensate for this undesirable effect with the use of fixed magnets at both ends of the array, and ensure that the frequencies across the array are as uniform as possible.   

In order to measure the local magneto-mechanical susceptibility ({a}ngular displacement per applied torque) of the array, we employ a frequency domain forced-response measurement. We harmonically drive each resonator with a small solenoid coil, and measure the resulting torsional oscillation as function of frequency using a Hall sensor placed nearby (Fig.~\ref{fig:1dResults}a). {The susceptibility spectrum is then calculated as the ratio of the angular amplitude to the applied torque at a given frequency. This susceptibility spectrum measured at each resonator is directly proportional to the local spectral density of states. By averaging these local measurements from all the resonators in the array we obtain the system-wide density of states, up to a proportionality factor.}

\vspace{12pt} 

The resonator arrays studied in this paper are characterized by mechanical second-order equations of motion. 
Using a slowly varying envelope approximation 
\cite{Supplement}
we rewrite these equations to obtain the momentum space Bloch Hamiltonian. For a 1D SSH chain 
(Fig. \ref{fig:Concept}a) composed of magneto-mechanical resonators, having intra-cell coupling $\gamma_1$ and inter-cell coupling $\gamma_2$ this Hamiltonian reads as follows:
\begin{eqnarray}\begin{aligned}
	\medmuskip=2mu
	H(k_x) = \begin{bmatrix}
	\frac{\omega^2-\omega_r^2}{2\omega}-\frac{ic}{2I}-\frac{c}{2I\omega} & \frac{\gamma_2}{\omega}e^{-ik_x}+\frac{\gamma_1}{\omega} \\
	\frac{\gamma_2}{\omega}e^{ik_x}+\frac{\gamma_1}{\omega} & \frac{\omega^2-\omega_r^2}{2\omega}-\frac{ic}{2I}-\frac{c}{2I\omega}
	\end{bmatrix}.
	\end{aligned}
	\label{eq:Hamiltonian}
\end{eqnarray}
Here $k_x$ is the momentum, $c$ is a viscous damping coefficient, $I$ is the mechanical moment of inertia, $\omega_r$ is the effective resonance frequency (including both the mechanical resonance frequency and magnetostatic spring effect \cite{grinberg2019magnetostatic}), and $\omega$ is the angular frequency of an harmonically oscillating solution. 
Note that the Hamiltonian in Eq. \ref{eq:Hamiltonian} differs from an SSH model Hamiltonian \cite{su1980soliton} only by a term proportional to the identity matrix, therefore, the eigenstates of the two are identical, and thus, so are all of the topological properties.

We experimentally assembled
 1D SSH chains 
 comprised from four unit cells periodically arranged along the $\hat{x}$ direction. We demonstrate the three configurations of the SSH model corresponding to topologically trivial ($\gamma_1>\gamma_2$), critical ($\gamma_1=\gamma_2$), and nontrivial ($\gamma_1<\gamma_2$) phases by adjusting the relative distances between the resonators, i.e. {changing the ratio between the inter-cell and intra-cell couplings $\gamma_2/\gamma_1$}. 
The measured system-wide normalized mechanical susceptibility (mechanical density of states) of these configurations is presented in Fig. \ref{fig:1dResults}b-d where different response bands are highlighted. The spatial mode distribution corresponding to the highlighted regions is presented as well.
For an array in the topologically trivial phase (Fig. \ref{fig:1dResults}b) two bands separated by a band-gap were obtained as expected \cite{su1980soliton}, and the states in both bands are spatially distributed almost evenly across the array. 
Since the magnetic coupling between resonators decays cubically with distance, next-nearest neighbor coupling is inherent in the system, which breaks the sub-lattice (chiral) symmetry. {This manifests as a slight asymmetry in the height and spectral width of the two bands, even though the number of states in them is the same}. 
For the critical configuration where the band gap is closed (Fig. \ref{fig:1dResults}c), we observe a single band with almost uniform spatial distribution of states across the array. 
In the topologically nontrivial phase (Fig. \ref{fig:1dResults}d) three bands are identified. The lower and upper bands exhibit bulk modes, while the mid-gap band exhibits the edge localized modes as can be observed clearly by the spatial distribution plots.

To rule out the possibility of the mid-gap modes arising from geometrical edge properties in our structure, we produced an additional arrangement having a domain wall between a topologically trivial and nontrivial phases. 
This arrangement (Fig. \ref{fig:1dResults}e) is comprised of 4 unit cells, two of which are in the trivial phase and two in the nontrivial phase. { The region between} these two phases forms a domain wall and is {therefore} expected to exhibit a localized mid gap mode.
The average mechanical susceptibility measured in this system reveals four bands. The two outer bands are again observed to correspond to bulk modes, while the two states in the middle correspond to the localized modes formed at the boundaries of the topologically nontrivial phase.
The frequency of the localized mode at the domain wall is slightly shifted (compared to the frequency of the mode at the right edge) due to slight differences in the local magnetostatic spring effect.

\begin{figure}[!hp]
		\makebox[\textwidth][c]{\includegraphics[width=0.8\textwidth]{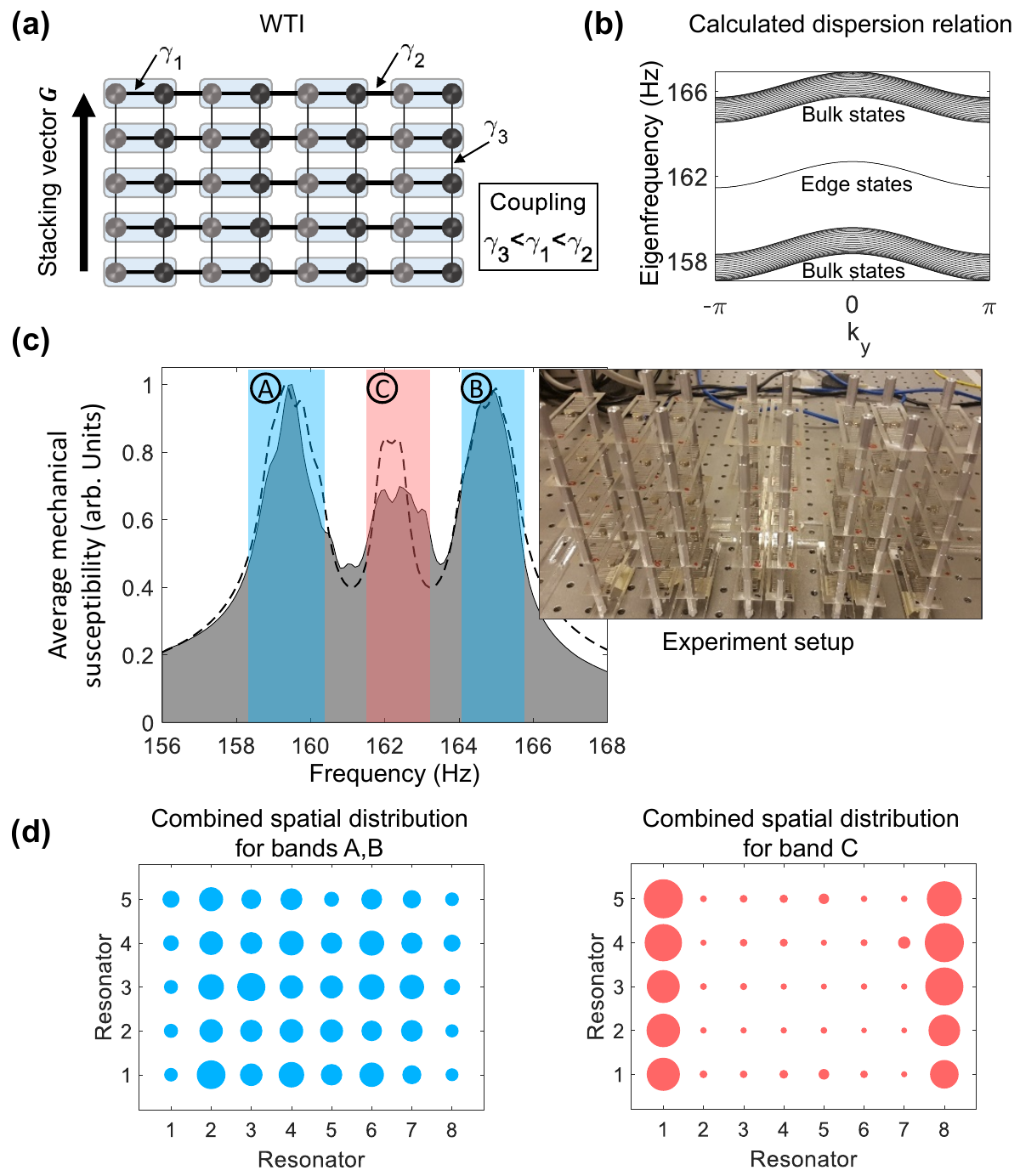}}
		\centering
		\caption{
			\textbf{(a)} Schematic of a 2D WTI made from $\hat{x}$-oriented 1D SSH chains that are stacked along the $\hat{y}$ direction. The width of the coupling lines is illustrative of coupling strength between resonators in the array ($\gamma_3<\gamma_1<\gamma_2$).
			\textbf{(b)} Dispersion relation calculated for a configuration similar to (a) but with 50 resonators (25 unit cells) in $\hat{x}$ and periodic boundary condition in $\hat{y}$, shows the existence of mid-gap modes. 
			\textbf{(c)} Plot of the experimentally measured  system-wide average mechanical susceptibility (normalized) versus frequency. The dashed line is the theoretical model prediction. We identify three bands of interest which are highlighted. The inset shows a photograph of the experimental setup.
			\textbf{(d)} Spatially resolved mode distribution (circle size corresponds to excitation amplitude) over the bands identified in (c).  For each resonator the measured susceptibility is averaged over the highlighted frequency region and the circle size corresponds to the averaged magnitude. Since the response characteristics of bands A,B {are} similar we combine their contributions. The lower and upper bands (A, B, blue in subfigure (c)) correspond to bulk states, while the middle band (C, pink in subfigure (c)) corresponds to the states formed on edges parallel to stacking vector ${\bf G}$.
			}
		\label{fig:2dResults}
\end{figure}

\vspace{12pt}


We now stack multiple SSH chains in the nontrivial phase (Fig. \ref{fig:1dResults}d), to produce a 2D WTI (as illustrated in Fig. \ref{fig:2dResults}a)
 with a  stacking (reciprocal) vector ${\bf G} = 2\pi\hat{y}/d_y$ where $d_y$ is the distance between layers in the $\hat{y}$ direction which we set to unity for convenience.  
The 1D TI chains are weakly and uniformly coupled along the vertical direction with coupling rate $\gamma_3$ (in our experiments we set the couplings $\gamma_3<\gamma_1<\gamma_2$). 
\begin{eqnarray}\begin{aligned}
\label{eq:H_WTI}
	\medmuskip=-3mu
	\thinmuskip=-2mu
	\thickmuskip=-2mu
	\nulldelimiterspace=0pt
	\scriptspace=0pt
	H\,(k_x,\, k_y)\smallequal\,\begin{bmatrix}
	\frac{\omega^2-\omega_r^2}{2\omega}\smallminus\frac{ic}{2I}\smallminus\frac{c}{2I\omega}\smallplus\frac{2\gamma_3}{\omega} \smallcos & \frac{\gamma_2}{\omega}e^{\smallminus ik_x}\smallplus\frac{\gamma_1}{\omega} \\
	\frac{\gamma_2}{\omega}e^{ik_x}\smallplus\frac{\gamma_1}{\omega} & \frac{\omega^2-\omega_r^2}{2\omega}\smallminus\frac{ic}{2I}\smallminus\frac{c}{2I\omega}\smallplus\frac{2\gamma_3}{\omega} \smallcos
	\end{bmatrix}.
\end{aligned}\end{eqnarray}
where $k_{x,y}$ are {momenta} along $\hat{x},\hat{y}$ directions respectively. 
\cite{Supplement}
In Fig.~\ref{fig:2dResults}b we evaluate the dispersion relation of the 2D WTI, with periodic boundary conditions in the $\hat{y}$ direction, and open boundaries in the $\hat{x}$ direction, showing the existence of the mid-gap modes\cite{franz2013topological}.
Due to the aforementioned anisotropy of a WTI, these mid-gap modes will not appear if the $\hat{x}$ direction is periodic and the $\hat{y}$ direction is open.

The measured average mechanical susceptibility of the entire 2D array (Fig.~\ref{fig:2dResults}c) reveals three bands. The lower and upper bands correspond to bulk modes (Fig.~\ref{fig:2dResults}d-left), and the middle band corresponds to localized modes on the left and right edges (Fig.~\ref{fig:2dResults}d-right) as expected. The individual susceptibility measurements of all 40 resonators are available in 
\cite{Supplement}.

\begin{figure}[!hp]
		\makebox[\textwidth][c]{\includegraphics[width=0.8\textwidth]{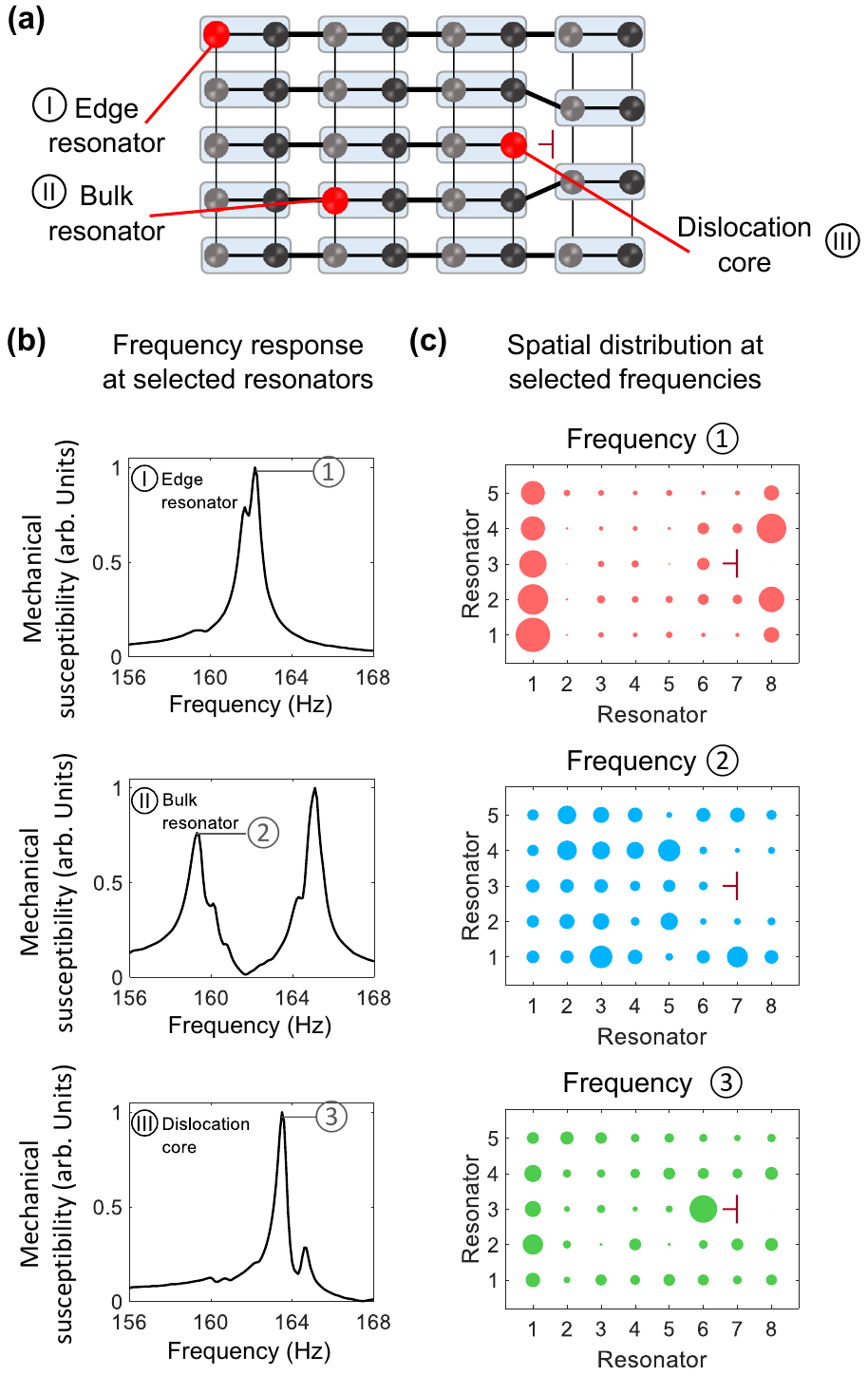}}
		\centering
		\caption{
			Observation of a trapped mode at a dislocation defect in a 2D WTI.
			\textbf{(a)} Illustration of the 2D WTI composed of 1D TI chains (dimerized coupling) along $\hat{x}$, and approximately uniform coupling in the $\hat{y}$ direction. A unit cell was removed from the right edge of the middle row to create the dislocation defect, and the $\hat{y}$-axis positions of the last column were adjusted to compensate. Three resonators of interest (red dots) are identified.
			\textbf{(b)} The measured susceptibilities at the resonators of interest show the array characteristics and are used to identify frequencies that may correspond to states of interest. 
			\textbf{(c)} Spatial resolved distribution of states at the three identified frequencies are presented, confirming the edge mode, the bulk modes, and the trapped mode at the dislocation defect.
			}
		\label{fig:2dResultsDefect}
\end{figure}

\vspace{12pt}


Next, we test the occurrence of  0D mid-gap modes trapped at the core of a topological defect. 
For this, we produce a dislocation with Burgers' vector ${\bf B} = \hat{y}$ by terminating one of the TI chains early as shown in Fig.~\ref{fig:2dResultsDefect}a. 
The $\hat{x}$ axis positions of the resonators in the last column are then adjusted to restore translation symmetry and to roughly restore an even inter-chain coupling $\gamma_3$.
In this configuration we expect localized modes on the left and right edges as in the case of the WTI without the dislocation, and an additional mode trapped at the dislocation core (Fig.~\ref{fig:Concept}c).
We choose three resonators as representative of the array characteristics: (I) a resonator on the left edge that is expected to have a localized mode, (II) a resonator deep inside the array that is expected to show the bulk band gap, and (III) the dislocation core which is also expected to show a localized mode. 
The measured normalized mechanical susceptibility of these three resonators is presented in Fig.~\ref{fig:2dResultsDefect}b, and matches the theoretical predictions. Once again, the individual susceptibility measurements of all 38 resonators are provided in the 
\cite{Supplement}. 
These measurements allow us to identify frequencies of interest that should correspond to the edge mode, the bulk bands, and the dislocation defect mode.
In Fig.~\ref{fig:2dResultsDefect}c we present the spatial distribution of states over the entire system of 38 resonators at the identified frequencies, using the  system-wide measured susceptibility. This spatial visualization confirms that the bulk and edge modes all appear as anticipated, and that the dislocation core now harbors a single trapped mode. 

We note that the magneto-mechanical implementation shown here is inherently disordered, e.g. resonators frequencies can only be matched within $\pm$0.1 Hz, which is significant compared to $\approx 2.5$ Hz bandgap. The cubic dependence of magneto-mechanical coupling makes the array very sensitive to variations in inter-resonator distance, and susceptible to next-nearest neighbor coupling. In spite of this, we observe that the governing topological characteristics for WTIs still hold.


\section{Discussion}

{
In this work we have presented an experimental demonstration of the anisotropic edge
modes of a WTI, as well as a trapped mode at the core of a dislocation
defect. These results confirm the ability of WTIs to generate both 1D and 0D topological 
bound states from the same 2D bulk system, which are the remarkable implications of a topological index theorem. 
	Broadly speaking, it has been a common assumption that topological insulators protect states that are one dimension lower than that of the host material. This assumption has came to be challenged recently by the rise of high-order topological insulators \cite{peterson2018quantized,benalcazar2017electric,benalcazar2017quantized,imhof2018topolectrical,serra2018observation,noh2018topological,ni2019observation,xue2019realization}, that exhibits protected states of higher co-dimension. Our work experimentally demonstrates yet another path to obtain lower dimensional protected states,  without relying on high-order topology. The expanded capability of producing both 0D and 1D states within the same 2D material could open up new avenues for robust systems, such as sensors, filters, and  other  signal processing devices  that are resilient to disorder that might appear in manufacturing  or during operation.}

\section*{Acknowledgments}

We acknowledge funding support from the National Science Foundation
Emerging Frontiers in Research and Innovation NewLAW program (grant
EFMA-1641084), the Office of Naval Research Director of Research Early
Career Grant (grant N00014-16-1-2830) and Multidisciplinary University Research Initiative (grant N00014-20-1-2325). This work was supported in part by the Zuckerman STEM Leadership Program.

\nocite{apsrev41Control}
\bibliographystyle{apsrev4-1}
\bibliography{Weak_TI_refs}

\end{document}


\title{\textbf{Supplementary material: Trapped state at a dislocation in a weak magnetomechanical topological
   insulator}}

\author{Inbar Hotzen Grinberg$^1$, Mao Lin$^2$,
	\mbox{Wladimir A. Benalcazar$^2$},
	\mbox{Taylor L. Hughes$^{2\ast}$}, and Gaurav Bahl$^{1\ast}$\\
	\footnotesize{$^1$ Department of Mechanical Science and Engineering, $^2$ Department of Physics,}\\
	\footnotesize{University of Illinois at Urbana-Champaign, Urbana, Illinois 61801, USA}\\
	\footnotesize{$^\ast$ To whom correspondence should be addressed; hughest@illinois.edu, bahl@illinois.edu} 
}

\date{}

	\vspace*{-2cm}
	{\let\newpage\relax\maketitle}
	
\beginsupplement

\section{Equations of motion and derivation of system Hamiltonian}
\label{sec:Equations of motion and derivation of system Hamiltonian}

In this section we will derive the equations of motion of the magneto-mechanical resonator array, and using slowly varying envelope approximation (SVEA) derive the Hamiltonian shown in Eq. 1 of the main text.
A more comprehensive derivation can be found in the supplementary material of \cite{grinberg2019robust}.

\begin{figure}[t]
\centering
\includegraphics[width=0.5\columnwidth]{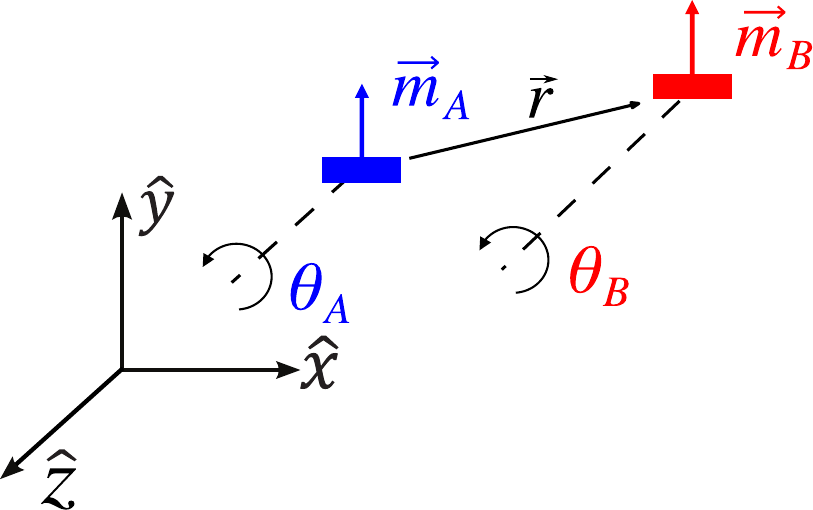}
\caption{{We illustrate two magneto-mechanical resonators A,B with relative position $\vec{r}$. The resonators each have a single rotational degree of freedom around the $\hat{z}$ axis given by $\theta_A$ and $\theta_B$. The magnetic dipole moments are initially oriented along the $\hat{y}$ axis.}}
\label{fig:TwoMagnets}
\end{figure}

{
In Fig.~\ref{fig:TwoMagnets}, we present the simple case of two interacting magneto-mechanical resonators. We consider each resonator as a point dipole pointing in the $\hat{y}$ direction at rest, with a single rotational degree of freedom around the $\hat{z}$ axis. This assumption is heuristically acceptable as long as the distance between magnets is greater than their largest geometrical dimension. Let us now consider two such resonators A and B at a relative position 
{$\vec{r}=r(u\hat{x}+v\hat{y}+w\hat{z})$ as shown in Fig.~\ref{fig:TwoMagnets}, where $u \hat{x} + v \hat{y} + w \hat{z}$ is the unit vector indicating the direction from A to B, and $r$ is the distance. }
The { $\hat{z}$-directed} torque acting on  dipole  $A$ due to dipole  $B$, after application of the small angle approximation, is given by \cite{grinberg2019magnetostatic}
\begin{equation}
\tau_{AB}(\theta_A,\theta_B)=\frac{\mu_0m^2}{4 \pi r^3 }((u^2-2v^2)\theta_A+(2u^2-v^2)\theta_B).
\label{eq:torque}
\end{equation}
Here $\mu_0=4\pi\times10^\text{-7}$ $N/A^2$ is the magnetic permeability of free space, $m=|\vec{m}_A|=|\vec{m}_B|$ is the magnetic dipole moment, 
and $\theta_A, \theta_B$ are the angular displacements 
for $A, B$ respectively, described using right-hand convention along the $\hat{z}$ axis as shown in Fig.~\ref{fig:TwoMagnets}.
}
From the perspective of resonator $A$, the dependence of this torque on $\theta_A$ implies a spring-like effect, while the dependence on $\theta_B$ implies a coupling to the adjacent resonator $B$. The magnitude of the magnetic torque is proportional to {$1/r^{3}$}, {implying that the coupling} can be controlled by adjusting the spacing between resonators. 

Our 1D {arrays are} built from multiple copies of the magneto-mechanical resonator,  spaced along $\hat{x}$, i.e. {$\vec{r}={r}\hat{x}$}. In this case the torque simplifies to 
\begin{equation}
\tau_\text{AB}(\theta_A,\theta_B)=\gamma \left( \frac{1}{2}\theta_A+\theta_B \right)
\label{eq:torque_2}
\end{equation}
where
\begin{equation}
\gamma=\frac{\mu_0m^2}{2 \pi {r}^3}
\label{eq:gamma}
\end{equation}
is the coupling coefficient as well as the spring softening coefficient{\cite{grinberg2019magnetostatic}}. 

\begin{figure}[t]
\centering
\includegraphics[width=0.5\columnwidth]{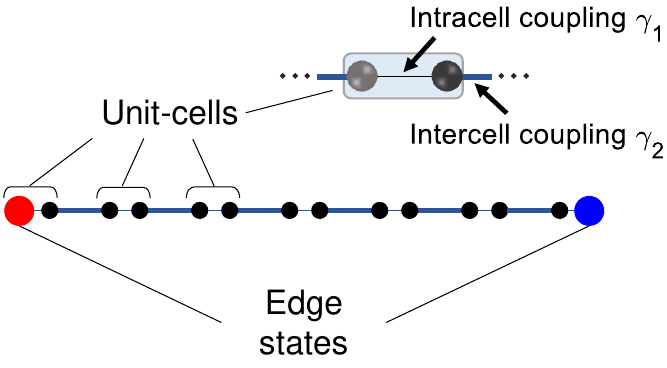}
\caption{{Illustration of a dimerized  array forming a 1D topological insulator - the SSH chain. For visual simplicity we illustrate the limiting case where intra-cell coupling {$\gamma_1$ is zero}. This 1D array exhibits two mid-gap states localized on the two edges.}
}
\label{fig:S2}
\end{figure}

{One unit cell of our typical test structures is described}  as a pair of adjacent resonators with intra-cell coupling $\gamma_1$. The inter-cell coupling between adjacent unit cells is denoted by $\gamma_2$. {We depict such a 1D array in Fig.~\ref{fig:S2}}. {The} equations of motion describing an infinite 1D array of such unit cells (assuming only nearest neighbor coupling) {can be} written as
\begin{equation}
\medmuskip=0mu
\begin{cases}
\begin{array}{l}
\label{eq:EOM_magneto}			
\ddot{\theta}_{n,A} + \frac{c}{I}\dot{\theta}_{n,A} + \omega_r^2\theta_{n,A} - \gamma_2 \theta_{n-1,B} - \gamma_1 \theta_{n,B} = 0  \\
\ddot{\theta}_{n,B} + \frac{c}{I}\dot{\theta}_{n,B} + \omega_r^2\theta_{n,B} - \gamma_2 \theta_{n+1,A} - \gamma_1  \theta_{n,A} = 0 
\end{array}
\end{cases}	
\end{equation}	
where $n$ denotes the index of the unit cell, $I$ is the cross sectional moment of inertia of a single resonator, and $c$ is the viscous damping coefficient. We define the effective resonance frequency $\omega_r^2=\omega_0^2 - \gamma_1 - \gamma_2$, where $\omega_0^2=\kappa/I$ is the natural mechanical resonance frequency, and $\kappa$ is the torsional spring constant arising from mechanics alone. 

We can now assume a time-harmonic solution and apply the slowly varying envelop approximation (SVEA) to the above equations. This is acceptable as long as any transients in the envelope of the harmonic oscillation, which is associated with resonance decay time, change at a much longer timescale than the oscillation time period. This difference {between transients and the oscillation period} is about two orders of magnitude in our system and therefore the SVEA is a permissible simplification.

We {continue} by assuming an oscillating solution of the form
\begin{eqnarray}\begin{aligned}
\label{eq:SVAA1}
\theta_{n,\eta}(t) = x_{n,\eta}(t) e^{i\omega t} +c.c. ~, 
\end{aligned}\end{eqnarray}
where $\omega$ is the frequency of the external drive and $x_{n,\eta}(t)$ is the complex-valued amplitude of oscillation {for site $\eta=A,B$ in the $n^{\text{th}}$ unit cell}. 
The time derivatives of the assumed solution are given by
\begin{eqnarray}\begin{aligned}
\label{eq:SVAA2}
\dot{\theta}_{n,\eta}(t) &= \dot{x}_{n,\eta}(t) e^{i\omega t} + i\omega x_{n,\eta}(t) e^{i\omega t} +c.c. ~,
\end{aligned}\end{eqnarray}
and
\begin{eqnarray}\begin{aligned}
\label{eq:SVAA3}
\ddot{\theta}_{n,\eta}(t) &= \ddot{x}_{n,\eta}(t) e^{i\omega t} + 2i\omega \dot{x}_{n,\eta}(t) e^{i\omega t} -\omega^2x_{n,\eta}(t) e^{i\omega t} +c.c.  ~.
\end{aligned}\end{eqnarray}
The SVEA {implies} that the amplitude $x_{n,\eta}(t)$ varies {slowly} and therefore we can set $\ddot{x}_{n,\eta}(t)=0$. 
Substituting Eq.~\eqref{eq:SVAA1}-\eqref{eq:SVAA3} into Eq.~\eqref{eq:EOM_magneto} and rearranging we get
\begin{eqnarray}\begin{aligned}
	\left(i+\frac{c}{2I\omega}\right)\dot{x}_{n,A} &= \left(\frac{\omega^2-\omega_0^2+\gamma_1+\gamma_2}{2\omega}-\frac{ic}{2I}\right)x_{n,A} + \frac{\gamma_2 }{\omega}x_{n-1,B} + \frac{\gamma_1 }{\omega} x_{n,B} \\
	\left(i+\frac{c}{2I\omega}\right)\dot{x}_{n,B} &= \left(\frac{\omega^2-\omega_0^2+\gamma_1+\gamma_2}{2\omega}-\frac{ic}{2I}\right)x_{n,B} + \frac{\gamma_2 }{\omega}x_{n+1,A}  + \frac{\gamma_1 }{\omega} x_{n,A}.
\label{eq:MotionAfterSVEA}
\end{aligned}\end{eqnarray}
Next, to obtain the Bloch Hamiltonian, we take the following Fourier transformation
\begin{eqnarray}\begin{aligned}
x_{n,\eta}(t) = \frac{1}{\sqrt{N}}\sum_k e^{-ink}x_{k,\eta}(t),
\end{aligned}\end{eqnarray}
where $1/\sqrt{N}$ is the normalization constant. Applying this to Eq. \ref{eq:MotionAfterSVEA} and writing it in matrix form yields {the relation}
\begin{align}
	\left(i+\frac{c}{2I\omega}\right)\frac{d}{dt}
\begin{bmatrix}
x_{k,A} \\
x_{k,B} 
\end{bmatrix}
= \begin{bmatrix}
\frac{\omega^2-\omega_0^2+\gamma_1+\gamma_2}{2\omega}-\frac{ic}{2I} & \frac{\gamma_2}{\omega}e^{-ik}+\frac{\gamma_1}{\omega} \\
\frac{\gamma_2}{\omega}e^{ik}+\frac{\gamma_1}{\omega} & \frac{\omega^2-\omega_0^2+\gamma_1+\gamma_2}{2\omega}-\frac{ic}{2I}
\end{bmatrix}
\begin{bmatrix}
x_{k,A} \\
x_{k,B} 
\end{bmatrix},
\end{align}
where the Bloch Hamiltonian is
\begin{eqnarray}\begin{aligned}
H(k) = \begin{bmatrix}
\frac{\omega^2-\omega_0^2+\gamma_1+\gamma_2}{2\omega}-\frac{ic}{2I}-\frac{c}{2I\omega} & \frac{\gamma_2}{\omega}e^{-ik}+\frac{\gamma_1}{\omega} \\
\frac{\gamma_2}{\omega}e^{ik}+\frac{\gamma_1}{\omega} & \frac{\omega^2-\omega_0^2+\gamma_1+\gamma_2}{2\omega}-\frac{ic}{2I}-\frac{c}{2I\omega}
\end{bmatrix}.
\end{aligned}\end{eqnarray}
Subtituting $\omega_r^2=\omega_0^2-\gamma_1-\gamma_2$ we reproduce Eq. 1 of the main text:

\begin{eqnarray}\begin{aligned}
H(k) = \begin{bmatrix}
\frac{\omega^2-\omega_r^2}{2\omega}-\frac{ic}{2I}-\frac{c}{2I\omega} & \frac{\gamma_2}{\omega}e^{-ik}+\frac{\gamma_1}{\omega} \\
\frac{\gamma_2}{\omega}e^{ik}+\frac{\gamma_1}{\omega} & \frac{\omega^2-\omega_r^2}{2\omega}-\frac{ic}{2I}-\frac{c}{2I\omega}
\end{bmatrix}.
\end{aligned}\end{eqnarray}

\section{Derivation of the 2D array Hamiltonian}

\begin{figure}[t]
\centering
\includegraphics[width=0.6\columnwidth]{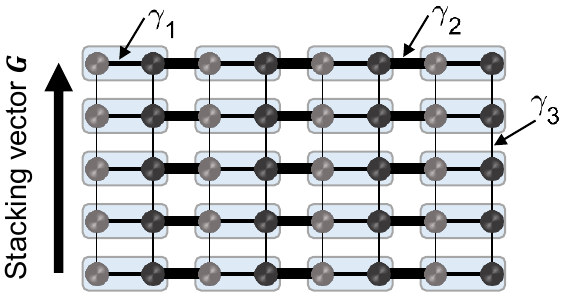}
\caption{Schematic of a 2D WTI made from $\hat{x}$-oriented 1D SSH chains that are stacked along the $\hat{y}$ direction. The { thickness} of the coupling lines is illustrative of coupling strength between resonators in the array ($\gamma_3<\gamma_1<\gamma_2$)}
\label{fig:S3}
\end{figure}

In this section, we shall derive the Bloch Hamiltonian of {an} infinite 2D array, which consists of a stack of 1D array{s} { as previously } described in Sec.~\ref{sec:Equations of motion and derivation of system Hamiltonian}. 
{
We depict such a 2D array in Fig.~\ref{fig:S3}.
} 
The coupling along the $\hat{y}$-direction, denoted as $\gamma_3$, is taken to be the same for all unit cells, such that the equation of motion of {the} 2D array reads
\begin{equation}
\medmuskip=0mu
\begin{cases}
\begin{array}{l}
\label{eq:EOM_magneto_2D}			
\ddot{\theta}_{n,m,A} + \frac{c}{I}\dot{\theta}_{n,m,A} + \omega_r^2\theta_{n,m,A} - \gamma_2 \theta_{n-1,m,B} - \gamma_1 \theta_{n,m,B} - \gamma_3(\theta_{n,m-1,A}+\theta_{n,m+1,A})= 0,  \\
\ddot{\theta}_{n,m,B} + \frac{c}{I}\dot{\theta}_{n,m,B} + \omega_r^2\theta_{n,m,B} - \gamma_2 \theta_{n+1,m,A} - \gamma_1  \theta_{n,m,A} - \gamma_3(\theta_{n,m-1,B}+\theta_{n,m+1,B})= 0.
\end{array}
\end{cases}	
\end{equation}	
Here, the {location of each } unit cell is labelled {with indices} $(n,m)$ and consists of two {resonators} labeled by $A,B$. We perform the same SVEA to the equation of motion, with  similar assumptions
\begin{eqnarray}\begin{aligned}
\label{eq:SVAA1_2D}
\theta_{n,m,\eta}(t) = x_{n,m,\eta}(t) e^{i\omega t}+c.c. ~.
\end{aligned}\end{eqnarray}
Upon discarding the second order time-derivative $\ddot{x}_{n,m,\eta}(t)$, 
 The resulting equation reads
\begin{eqnarray}\begin{aligned}
	\left(i+\frac{c}{2I\omega}\right)\dot{x}_{n,m,A} &= \left(\frac{\omega^2-\omega_0^2+\gamma_1+\gamma_2}{2\omega}-\frac{ic}{2I}\right)x_{n,m,A} + \frac{\gamma_2 }{\omega}x_{n-1,m,B} \\
	&\quad+ \frac{\gamma_1 }{\omega} x_{n,m,B}+\frac{\gamma_3}{\omega}\left(x_{n,m-1,A}+x_{n,m+1,A} \right)\\
	\left(i+\frac{c}{2I\omega}\right)\dot{x}_{n,m,B} &= \left(\frac{\omega^2-\omega_0^2+\gamma_1+\gamma_2}{2\omega}-\frac{ic}{2I}\right)x_{n,m,B} + \frac{\gamma_2 }{\omega}x_{n+1,m,A}  \\
	&\quad+ \frac{\gamma_1 }{\omega} x_{n,m,A}+\frac{\gamma_3}{\omega}\left(x_{n,m-1,B}+x_{n,m+1,B} \right) .
\label{eq:MotionAfterSVEA_2D}
\end{aligned}\end{eqnarray}

Finally, upon applying the  Fourier transformation  ($k_{x,y}$ are {momenta} along $\hat{x}$ and $\hat{y}$ directions respectively) { to Eqn.~\ref{eq:MotionAfterSVEA_2D},}
\begin{eqnarray}\begin{aligned}
x_{n,m,\eta}(t) = \frac{1}{\sqrt{N}}\sum_{k_x,k_y} e^{-i(k_xn+k_ym)}x_{k_x,k_y,\eta}(t),
\end{aligned}\end{eqnarray}
 and writing the result in matrix form gives
\begin{eqnarray}\begin{aligned}
	i\frac{d}{dt}
\begin{bmatrix}
x_{k_x,k_y,A} \\
x_{k_x,k_y,B} 
\end{bmatrix}
= H(k_x,k_y)
\begin{bmatrix}
x_{k_x,k_y,A} \\
x_{k_x,k_y,B} 
\end{bmatrix},
\end{aligned}\end{eqnarray}
with the desired Bloch Hamiltonian
\begin{eqnarray}\begin{aligned}
\label{eq:H_WTI_app}
	\medmuskip=-3mu
	\thinmuskip=-2mu
	\thickmuskip=-2mu
	\nulldelimiterspace=0pt
	\scriptspace=0pt
	H\,(k_x,\, k_y)\smallequal\,\begin{bmatrix}
	\frac{\omega^2-\omega_r^2}{2\omega}\smallminus\frac{ic}{2I}\smallminus\frac{c}{2I\omega}\smallplus\frac{2\gamma_3}{\omega} \smallcos & \frac{\gamma_2}{\omega}e^{\smallminus ik_x}\smallplus\frac{\gamma_1}{\omega} \\
	\frac{\gamma_2}{\omega}e^{ik_x}\smallplus\frac{\gamma_1}{\omega} & \frac{\omega^2-\omega_r^2}{2\omega}\smallminus\frac{ic}{2I}\smallminus\frac{c}{2I\omega}\smallplus\frac{2\gamma_3}{\omega} \smallcos
	\end{bmatrix}.
\end{aligned}\end{eqnarray}
Here we have defined $\omega_r^2=\omega_0^2-\gamma_1-\gamma_2$ as in the 1D case.

\vspace{12pt}

\section{Derivation of the topological invariant for a 1D chain}
\label{sec:derivation_index}

In this section we derive the topological invariant of our 1D topological array. As discussed {above}, the Bloch Hamiltonian of the system is:
\begin{eqnarray}\begin{aligned}
\medmuskip=2mu
H(k) = \begin{bmatrix}
\frac{\omega^2-\omega_r^2}{2\omega}-\frac{ic}{2I}-\frac{c}{2I\omega} & \frac{\gamma_2}{\omega}e^{-ik}+\frac{\gamma_1}{\omega} \\
\frac{\gamma_2}{\omega}e^{ik}+\frac{\gamma_1}{\omega} & \frac{\omega^2-\omega_r^2}{2\omega}-\frac{ic}{2I}-\frac{c}{2I\omega}
\end{bmatrix}.
\end{aligned}
\end{eqnarray}
{This} Hamiltonian $H(k)$ differs from an typical SSH model Hamiltonian \cite{su1980soliton} only by a term proportional to an identity matrix, therefore, the eigenstates of the two are identical. Since the topological properties are completely determined by the eigenstates, we expect that our 1D topological array {will} share the same topological properties as the SSH model, namely that it is topologically nontrivial when $\gamma_1<\gamma_2$ and { is topologically trivial when} $\gamma_1>\gamma_2$. We shall demonstrate this explicitly in this section. 

We start by diagonalizing the Hamiltonian and pick its lowest eigenstate
\begin{align}
\label{eq:uk_app}
\ket{u(k)} = \frac{1}{\sqrt{2}}\begin{bmatrix}
\sqrt{\frac{\gamma_2e^{-ik}+\gamma_1}{\gamma_2e^{ik}+\gamma_1}} ~\\
1
\end{bmatrix},
\end{align}
which defines the topology of the system \cite{hasan2010colloquium,qi2011topological,moore2010birth}. Specifically, to conclude if the system is in the nontrivial or trivial phase, we have to calculate the Berry phase defined as \cite{bernevig2013topological}
\begin{align}
\label{eq:nu_app}
\nu=\frac{1}{2\pi} \int_{-\pi}^{\pi} \A(k) dk,
\end{align}
where $\A(k)=-i \matrixel{u(k)}{\partial_k}{u(k)}$ is the Berry connection \cite{xiao2010berry}. Upon plugging Eq.~\eqref{eq:uk_app} into Eq.~\eqref{eq:nu_app} we get
\begin{eqnarray}\begin{aligned}
\label{eq:nu_app_2}
\nu&=\frac{1}{2\pi} \int_{-\pi}^{\pi} \frac{-i}{2} \sqrt{\frac{\gamma_2e^{ik}+\gamma_1}{\gamma_2e^{-ik}+\gamma_1}}\partial_k\sqrt{\frac{\gamma_2e^{-ik}+\gamma_1}{\gamma_2e^{ik}+\gamma_1}}dk \\
&= \frac{1}{8\pi}\left[\text{Arg}\left(\frac{\gamma_2e^{-ik}+\gamma_1}{\gamma_2e^{ik}+\gamma_1}\big|_{k=\pi} \right)- \text{Arg}\left(\frac{\gamma_2e^{-ik}+\gamma_1}{\gamma_2e^{ik}+\gamma_1}\big|_{k=-\pi}\right) \right] \\
&= \left\{
\begin{array}{ll}
\frac{1}{2}, \quad \text{{modulo} } 1 \quad & \text{ when } \gamma_1<\gamma_2\\
0, \quad \text{{modulo} } 1 \quad &\text{ when } \gamma_1>\gamma_2
\end{array}
\right. .
\end{aligned}\end{eqnarray}
Here $\text{Arg}(z)$ is the argument of the complex number $z$. { $a=b\text{ modulo n}$ denotes that $a,b$ have the same reminder after division by n.} When {$\gamma_1<\gamma_2$ ($\gamma_1>\gamma_2$),} the topological array is in the nontrivial (trivial) phase with (without) edge modes at the end of the array. This is in agreement with the SSH model as expected \cite{su1980soliton}.

\vspace{12pt}

\section{Measured spectra of the 2D magneto-mechanical WTI}

\begin{figure}[!hp]
		\makebox[\textwidth][c]{\includegraphics[width=1.4\textwidth]{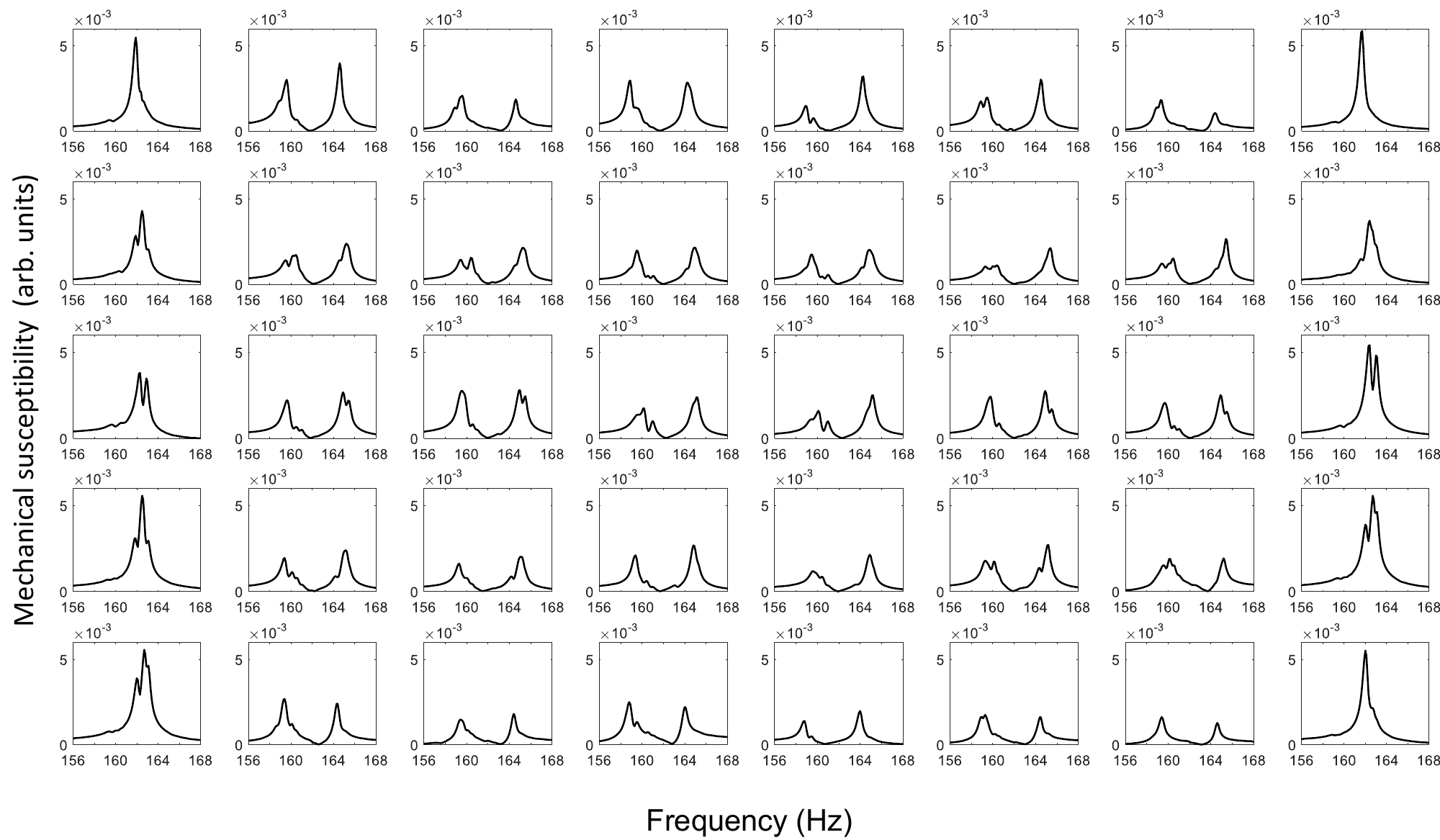}}
		\centering
		\caption{
				Measured susceptibility of  each resonator in the array, corresponding to Fig.~3 of the main text. As expected, the right and left edges host mid-gap localized modes while the bulk is gapped.
		}
	\label{fig:2dFull}
\end{figure}

\begin{figure}[!hp]
		\makebox[\textwidth][c]{\includegraphics[width=1.4\textwidth]{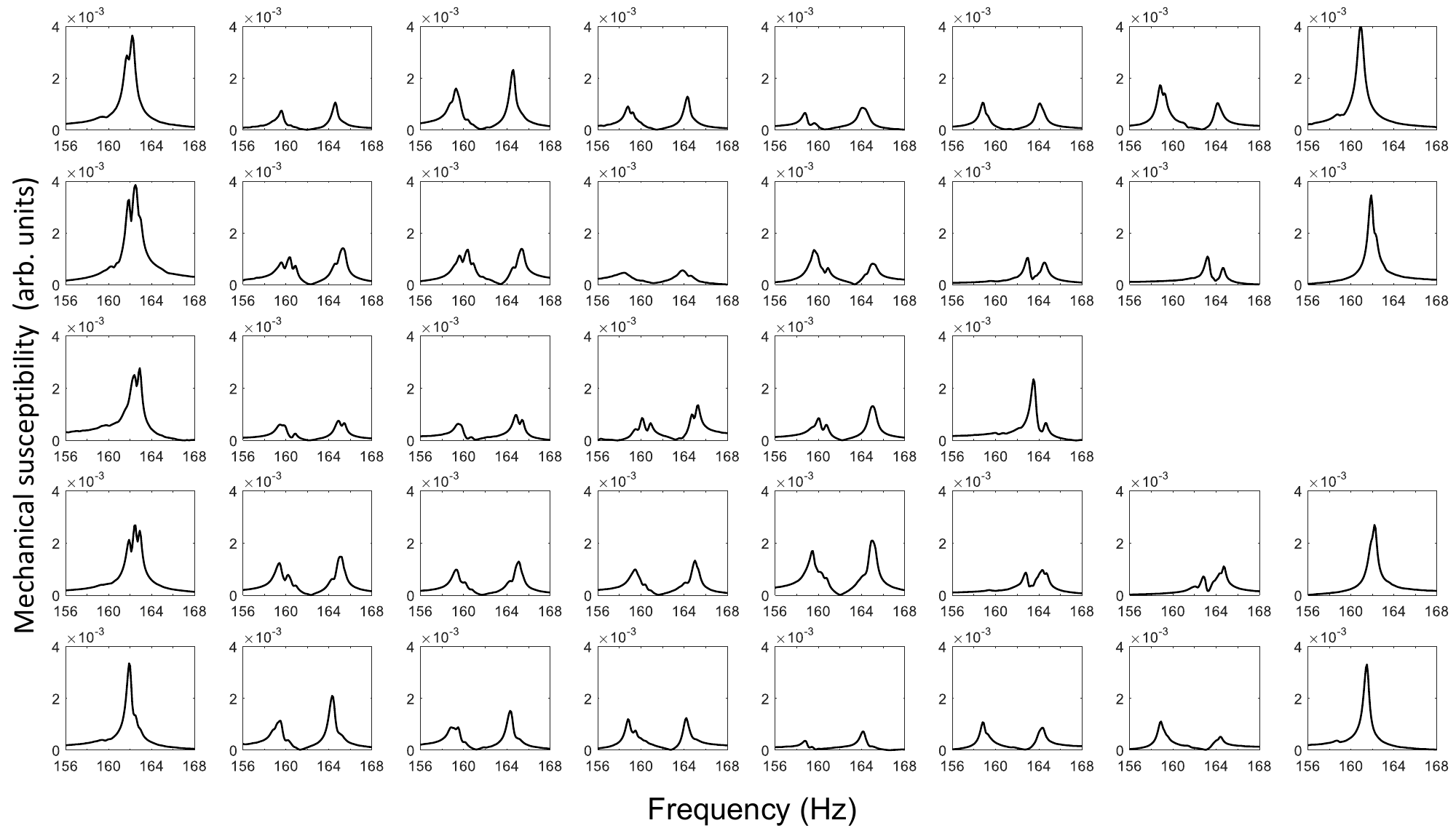}}
		\centering
		\caption{
				Measured susceptibility of each resonator in a WTI with a dislocation defect, corresponding to Fig.~4 of the main text. The left and right edges, as well as the dislocation core, host localized modes. The bulk of the array is gapped.
		}
	\label{fig:2dDefectFull}
\end{figure}

In this section we present the measured magneto-mechanical susceptibility of all resonators in the { array presented in the main text (Fig. \ref{fig:2dFull}), and for the same WTI with a dislocation defect (Fig. \ref{fig:2dDefectFull})}.  

\FloatBarrier

\nocite{apsrev41Control}
\bibliographystyle{apsrev4-1}
\bibliography{Weak_TI_refs}